\UseRawInputEncoding
\documentclass[showpacs,twocolumn,superscriptaddress,nofootinbib]{revtex4-1} 

\usepackage{hyperref}
\hypersetup{colorlinks=true,linkcolor=blue,citecolor=cyan}
\usepackage{graphicx}
\usepackage{xcolor, soul}
\sethlcolor{green}
\usepackage{dcolumn}
\usepackage{bm}
\usepackage{color}
\usepackage{enumitem}
\usepackage{mathrsfs}
\usepackage{amsmath}
\usepackage{amssymb}
\usepackage{cleveref}
\usepackage{orcidlink}

\crefname{equation}{Eq.}{Eqs.}
\crefname{figure}{Fig.}{Figs.}
\crefname{section}{Sec.}{Sec.}
\crefname{table}{Table}{Tables}
\usepackage{physics}

\setstcolor{red}

\newcommand{\etal}{\textit{et. al.~}}

\begin{document}

\title{
Efficiency of Penrose process 
in spacetime of axially symmetric 
magnetized Reissner-Nordstr\"{o}m black hole}

\author{Sanjar Shaymatov
\orcidlink{0000-0002-5229-7657}}
\email{sanjar@astrin.uz}

\affiliation{Institute for Theoretical Physics and Cosmology, Zheijiang University of Technology, Hangzhou 310023, China}
\affiliation{Akfa University,  Milliy Bog Street 264, Tashkent 111221, Uzbekistan}
\affiliation{Ulugh Beg Astronomical Institute, Astronomy Street 33, Tashkent 100052, Uzbekistan}
\affiliation{National University of Uzbekistan, Tashkent 100174, Uzbekistan}
\affiliation{Tashkent State Technical University, Tashkent 100095, Uzbekistan}

\author{Pankaj Sheoran
\orcidlink{0000-0001-8283-8744}}
\email{hukmipankaj@gmail.com}
\affiliation{Department of Physics, Gurukula Kangri (Deemed to be University), Haridwar 249 404, Uttarakhand, India}

\author{Ricardo Becerril
\orcidlink{0000-0001-9430-634X}}
\email{ricardo.becerril@umich.mx}
\affiliation{Instituto de F\'{\i}sica y Matem\'{a}ticas, Universidad Michoacana de San Nicol\'{a}s de Hidalgo,\\
Edificio C-3, 58040 Morelia, Michoac\'{a}n, M\'{e}xico}

\author{Ulises Nucamendi
\orcidlink{0000-0002-8995-7356}}
\email{unucamendi@gmail.com}
\affiliation{Instituto de F\'{\i}sica y Matem\'{a}ticas, Universidad Michoacana de San Nicol\'{a}s de Hidalgo,\\
Edificio C-3, 58040 Morelia, Michoac\'{a}n, M\'{e}xico}

\author{Bobomurat Ahmedov
\orcidlink{0000-0002-1232-610X}}
\email{ahmedov@astrin.uz}

\affiliation{Ulugh Beg Astronomical Institute, Astronomy St. 33, Tashkent 100052, Uzbekistan}
 \affiliation{National Research University TIIAME, Kori Niyoziy 39, Tashkent 100000, Uzbekistan}
\affiliation{National University of Uzbekistan, Tashkent 100174, Uzbekistan}

\date{\today}
\begin{abstract}
In this paper, we investigate the Penrose process in the purlieus of the axially symmetric magnetized Reissner-Nordstr\"{o}m black hole for both neutral and charged  particles. We start with the study of the geometry of the black hole and find the regions where the $g_{tt}$ component of the metric tensor is positive (i.e., $g_{tt}>0$). It is interestingly found that the condition $g_{tt}>0$ is fulfilled not only close to event horizon known as the ergosphere but also far 
away from the event horizon in the silhouette of potential wells. We also show that as the dimensionless magnetic field $B$ increases the silhouette of potential wells for which $g_{tt}>0$ grows correspondingly and eventually merges with the ergoregion when $B\gtrsim 1.6$.  Finally, we investigate the efficiency of the Penrose process for the axially symmetric magnetized black hole case and bring out the effect of magnetic field on it. Further, we also compare our results with the one for Kerr black hole.  
We show that when the charge $Q$ of the black hole is kept constant, the efficiency of the energy extraction process for the case of neutral particle (i.e., $q/m=0$)
first increases and then begins to decrease with rise in the value of $B$ field, in contrast to Kerr black hole where it always increases as the rotation parameter grows. 
However, for the case of charged particle (i.e., $q\neq 0$) the efficiency always increases with the rise in $B$ field and can go over $100\%$, when both $B$ and $q/m$ are large enough (say $B\approx1$ and $q/m>2.2$).
It is worth noting  that the existence of regions away from the horizon where $g_{tt}>0$ also favors the energy-extraction process away from effect of the black hole. However, the energy extraction from these regions is pure consequence of the magnetic field.

\end{abstract}
\maketitle

\section{Introduction}
\label{introduction}
In recent years, humankind has witnessed a slew of incredible discoveries, including the detection of gravitational waves from the merger of black holes \cite{LIGOScientific:2016aoc} and the first ever direct image of a black hole \cite{EventHorizonTelescope:2019dse}, thanks to modern instruments such as gravitational wave interferometers \cite{LIGOScientific:2007fwp,Accadia_2012} and the event horizon telescope (EHT) \cite{Fish:2016jil}. However, this is only the beginning of a new age in black hole physics, and in the future, with the help of more advanced new instruments \cite{KAGRA:2013rdx}, we will be able to make even more amazing discoveries and fine-tune our previous findings. 

In addition to gravitational waves and shadows of black holes, black holes are acknowledged as the most powerful sources of energy in nature. As a result, they are thought to create a tremendous electromagnetic environment in their surroundings, as well as being responsible for high-energy jet emission events capable of destroying nearby stars \cite{2011Nature:476} and galaxies. 


It is commonly accepted that a rotating black hole can function as an engine and be responsible for the high energy processes outlined above when wrapped in an external magnetic field, which is either imparted from a nearby star or created by the moment of nearby orbiting material. The electromagnetic Penrose process \cite{Wagh:1985vuj,1985JApA....6...85B,1986ApJ...307...38P,Wagh:1989zqa,Kolos:2020gdc}, also known as the magnetic Penrose process \cite{Dadhich:2018gmh} and the electric Penrose process \cite{1985JApA....6...85B,Tursunov21:EP}, is a high-energy emission event. The name comes from Roger Penrose's discovery of an energy extraction mechanism from a rotating black hole \cite{Penrose:1969pc} in the absence of any external fields. It has also been observed that in the absence of any external field, the maximum efficiency of the Penrose process for a spinning black hole (i.e., Kerr black hole) is around $20 \%$, which is purely due to the black hole's spin \cite{Toshmatov:2014qja}.
Contrary to this, superradiance in the electromagnetic environment can serve as the engine of ultrahigh energies \cite{Dadhich:2018gmh}, which is the far greater than any efficient mechanical engine. Therefore, it becomes important to study the magnetic Penrose process if one seeks the explanation of high-energy astrophysical phenomenon such as ultrahigh-energy cosmic ray \cite{PierreAuger:2017pzq,PierreAuger:2018qvk,Tursunov:2020juz}, relativistic jets \cite{Stuchlik:2015nlt,Istomin:2020oja} and particles observed with energy of about Gev ($\approx 10^2$) scale \cite{Ruffini:2018aiq,Kopacek:2018lgy}.

Recently, a detailed analysis of magnetic Penrose process under different regimes of efficiency, namely low, moderate and high, depending upon the magnetization and charging of a rotating black hole is presented in \cite{Tursunov:2019oiq}.

To be more specific about the mechanism, we are studying a process exemplified by (not only confined to) the disintegration of a master particle into two new particles $P_{E_{+}}$ and $P_{E_{-}}$, namely in the vicinity of a black hole.  It is assumed that the particle $P_{E_{+}}$ has a positive energy exceeding the energy of the master particle whereas the particle $P_{E_{-}}$ has negative energy. Even if the process may rely on electromagnetic interaction and not only limited to relativity of spacetime, the name of the process is used as electromagnetic Penrose process just for the sake of simplicity.

We explicitly find the region from which energy can be extracted from  axially symmetric magnetic Reissner-Nordstr\"{o}m black hole and show that the electromagnetic Penrose process is not limited only to the rotating black hole immersed in a magnetic field \cite{PhysRevD.29.2712,PhysRevD.30.1625}. There can be disconnected regions above and below the equatorial plane of the axially symmetric magnetic Reissner-Nordstr\"{o}m black hole known as toroidal region which is first reported by Gupta \etal \cite{Gupta:2021vww} for a rotating black hole immersed in a magnetic field. 

Here, in Sec.~\ref{Sec:Magnetized} we explore the geometry of magnetized Reissner-Nordstr\"om black hole spacetime. In
Sec.~\ref{Sec:Motion} we study the motion of charged particles in the strong gravitational and electromagnetic fields of the magnetized black hole. In Sec.~\ref{Sec:Penrose} we focus on the Penrose process in the vicinity of magnetized Reissner-Nordstr\"{o}m black hole. Sec.~\ref{Sec:conclusion} is devoted to the summary and conclusions. 
Unless otherwise mentioned, we adopt the spacetime metric signature $(-,+,+,+)$ and use the system of geometric units in which $G=c=1$ throughout the article.
\vspace{1cm}
\begin{figure}
 \centering
\includegraphics[scale=0.63]{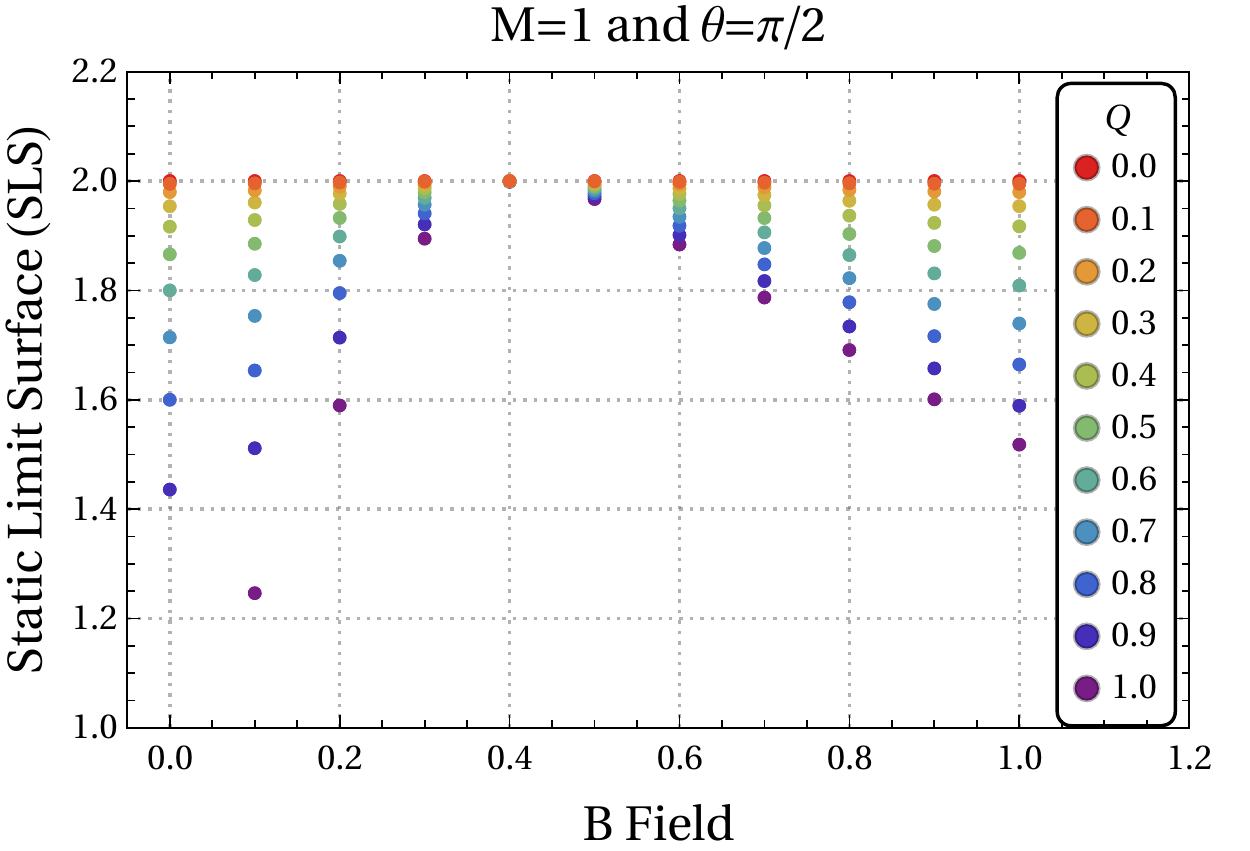}
\caption{\label{fig:SLS1} Plot shows the variation of outermost static limit surface as the function of $B$ for various combinations of $Q$ in the equatorial plane, $\theta=\pi/2$. Here, the mass parameters $M$ is set to unity. }
\end{figure}

\section{Magnetized Reissner-Nordstr\"{o}m black hole spacetime }\label{Sec:Magnetized}

The spacetime metric describing axially symmetric magnetized Reissner-Nordstr\"{o}m black hole in Schwarzschild coordinates ($t,r,\theta, \phi$) is given by \cite{Gibbons13}
\begin{eqnarray}\label{Eq:metric} d s^2 &=& H\, \left(-F dt^2 + F^{-1}\, dr^2  + r^2 d\theta^2\right) +
      H^{-1}\, r^2\sin^2\theta\, \nonumber\\
 &&\times\left(d\phi -\omega dt\right)^2\, ,
  \end{eqnarray}
where  
\begin{eqnarray}\label{Eq:fhw}
F&=& 1- \frac{2M}{r} + \frac{Q^2}{r^2}\, , \\
H &=& 1 +\frac{1}{2}B^2 (r^2\sin^2\theta + 3 Q^2\cos^2\theta)
\nonumber\\&&+
  \frac{1}{16} B^4 (r^2 \sin^2\theta + Q^2\cos^2\theta)^2\, ,\\
\omega &=& -\frac{2Q B}{r} + \frac{1}{2} Q B^3\, r (1+F
\cos^2\theta)\, , \end{eqnarray}
with parameters $M$ and $Q$ correspond to the black hole mass and charge.
Note that $B$ refers to the magnetic field parameter. The above metric reduces to the Reissner-Nordstr\"{o}m black hole one in the limit of $B\to 0$, while the Schwarzschild one in the limit of $B,Q\to 0$. 
It is worth noting that, interestingly, it turns out that the magnetized black hole also causes axially symmetric spacetime due to the existence of magnetic field  without any rotation. It is a remarkable property of magnetized Reissner-Nordstr\"{o}m black hole.  

Interestingly, the magnetized black hole's horizon has the form as 
\begin{eqnarray}
r_{\pm}=M\pm\sqrt{{M}^2-Q^{2}}\, ,
\end{eqnarray}
which is the same with the one for Reissner-Nordstr\"{o}m black hole. Apparently, the magnetic field parameter has no effect on the black hole horizon.

\begin{figure*}
\begin{tabular}{c c }
  \includegraphics[scale=0.7]{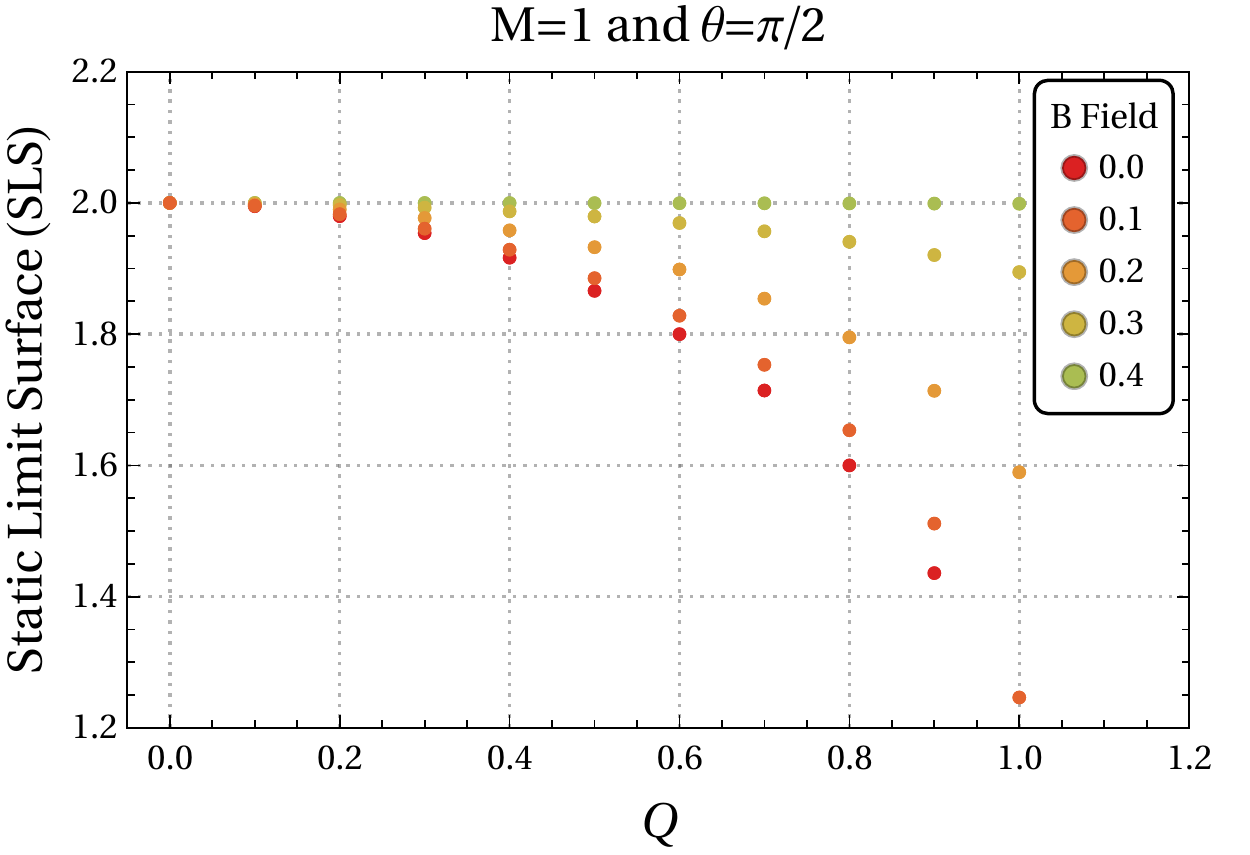}\hspace{-0.4cm}
  &  \includegraphics[scale=0.7]{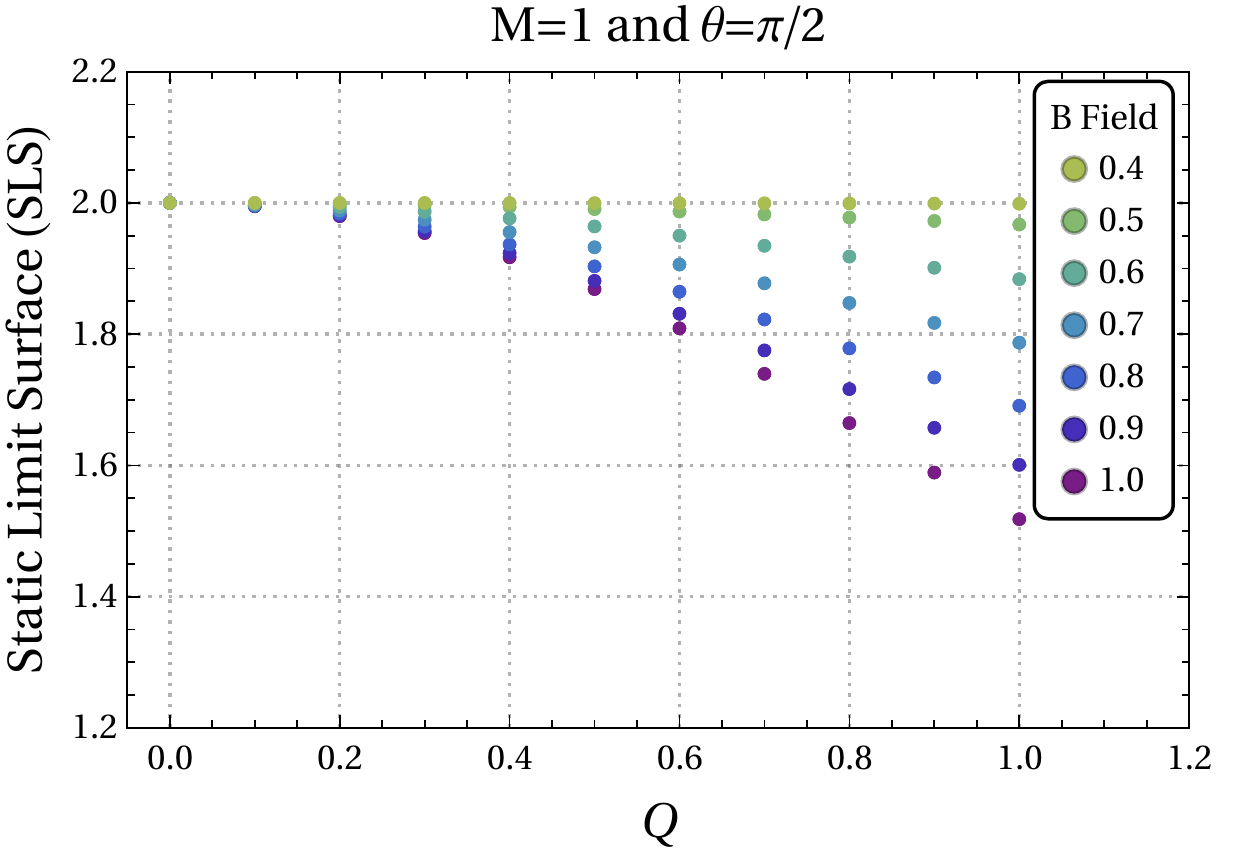}
\end{tabular}
	\caption{\label{fig:SLS2} 
In the equatorial plane, $\theta=\pi/2$, the variation of the outermost static limit surface as a function of black hole charge parameter $Q$ for various combinations of $B$ is shown. $B$ ranges from $0$ to $0.4$ in the {left panel} and from $0.4$ to $1.0$ in the {right panel}. Here, the mass parameter $M=1$.}
\end{figure*}

{Let us evaluate the total contraction of Ricci tensor (i.e., $ R_{\mu\nu}R^{\mu\nu}$) and also the Kretschmann scalar $\mathcal{K}$ to understand the property of the magnetized black hole singularity and compare it with the ones for Reissner-Nordstr\"{o}m and Schwarzschild cases. The scalar quantity $ R_{\mu\nu}R^{\mu\nu}$ for the magnetized black hole geometry reads as follows: 

\begin{widetext}
\begin{align}
    R_{\mu\nu}R^{\mu\nu}&=\left[\frac {1}{ \left( {B}^{4} \left( Q-r
 \right) ^{2} \left( Q+r \right) ^{2} \cos^{4}\theta
+2\,{B}^{2} \left(  \left( {r}^{2}{Q}^{2}-{r}^{4
} \right) {B}^{2}+12\,{Q}^{2}-4\,{r}^{2} \right) \cos^{2}\theta+ \left( {B}^{2}{r}^{2}+4 \right) ^{2}
 \right) ^{4}{r}^{8}}\right] 
 \nonumber\\
 &\times\Bigg[ 65536\,{B}^{8}{Q}^{4} \left( M{r}^{3}+\frac{{Q}^{4}}{8}-\frac{3{r}^{2}{Q}^{2}}{4}-\frac{3{r}^{4}}{8} \right)^{2} \cos^{8}\theta -131072\,{B}^{6}{Q}^{2} \Bigg( {r}
^{2}{Q}^{2} \left( M{r}^{3}+\frac{{Q}^{4}}{8}-\frac{3{r}^{2}{Q}^{2}}{4}-\frac{3{r}^{4}}{8}\right) 
\nonumber\\
&\left( Mr-\frac{{Q}^{2}}{4}-\frac{3{r}^{2}}{4}\right) {B}^{2}-\frac{{r}^{8}}{2}+\frac{5M{r}^{7}}{2}+ \left( -4\,{M}^{2}-{\frac {17\,{Q}^{2}}{8}}
 \right) {r}^{6}+8\,M{Q}^{2}{r}^{5}-{\frac {29\,{Q}^{4}{r}^{4}}{8}}-\frac{5
M{Q}^{4}{r}^{3}}{2}+{\frac {21\,{Q}^{6}{r}^{2}}{8}}
 \nonumber\\
 &-\frac{3{Q}^{8}}{8}
 \Bigg)\cos^{6}\theta+65536\,{B}^{
4} \Bigg( {r}^{4}{Q}^{4} \left( {\frac {15\,{r}^{4}}{32}}-\frac{5M{r}^{3
}}{4}+ \left( {M}^{2}+\frac{3{Q}^{2}}{16} \right) {r}^{2}-\frac{M{Q}^{2}r}{2}+{\frac 
{3\,{Q}^{4}}{32}} \right) {B}^{4}-16\, \left( {\frac {9\,{r}^{6}}{32}}
\right.
\nonumber\\
&\left.-{\frac {19\,M{r}^{5}}{16}}+ \left( {M}^{2}+{\frac {53\,{Q}^{2}}{64}}
 \right) {r}^{4}-M{Q}^{2}{r}^{3}-\frac{3{Q}^{4}{r}^{2}}{16}+{\frac {5\,M{Q}^
{4}r}{16}}-{\frac {3\,{Q}^{6}}{64}} \right) {r}^{2}{Q}^{2}{B}^{2}+
 \left( 16\,{M}^{2}+8\,{Q}^{2} \right) {r}^{6}
 \nonumber\\
 &-72\,M{Q}^{2}{r}^{5}+\frac{95Q^{4}r^{4}}{2}
+44\,M{Q}^{4}{r}^{3}-53\,{Q}^{6}{r}^{2}+
\frac{19{Q}^{8}}{2}\Bigg)\cos^{4}\theta 
-16384\,{B}^{2} \Bigg( {Q}^{4}{r}^{6} \left( Mr-\frac{{Q}^{2}}{4}-\frac{3{r}^
{2}}{4}\right) {B}^{6}
 \nonumber\\
&+ \left( -24\,{Q}^{2}{r}^{8}+60\,M{Q}^{2}{r}^{7}+
 \left( -32\,{M}^{2}{Q}^{2}-7\,{Q}^{4} \right) {r}^{6}+3\,{Q}^{6}{r}^{
4} \right) {B}^{4}+ \left( -64\,M{r}^{7}+ \left( 128\,{M}^{2}+176\,{Q}
^{2} \right) {r}^{6}\right.
\nonumber\\
&\left.-384\,M{Q}^{2}{r}^{5}+28\,{Q}^{4}{r}^{4}+176\,M{Q}
^{4}{r}^{3}-60\,{Q}^{6}{r}^{2} \right) {B}^{2}+64\,M{Q}^{2}{r}^{3}-48
\,{Q}^{6}+48\,{Q}^{4}{r}^{2}-64\,{Q}^{2}{r}^{4} \Bigg)\cos^{2}\theta
 \nonumber\\
 &+1024\, \left( {B}^{4}{Q}^{2}{r}^{
4}+ \left( 32\,M{r}^{3}-24\,{r}^{2}{Q}^{2}-16\,{r}^{4} \right) {B}^{2}
+16\,{Q}^{2} \right) ^{2} \Bigg]\, ,\label{eq:Ricci_sq} 
\end{align}
\end{widetext}
which in the limit $B\to 0$ recovers the Reissner-Nordstr\"{o}m case (i.e., $R_{\mu\nu}R^{\mu\nu}=4Q^{4}/r^{8}$) and in the limit $B,Q\to0$ reduces to the Schwarzschild case (i.e., $R_{\mu\nu}R^{\mu\nu}=0$). The explicit expression for $\mathcal{K}$ turns out to be very long and complicated. Therefore, we resort to simplify $\mathcal{K}$ as
\begin{widetext}
\begin{align}
    \mathcal{K}&=\frac {N(r,M,Q,B,\theta)}{ \left[ {B}^{4} \left( Q-r \right) ^{2}
 \left( Q+r \right) ^{2} \left( \cos \left( \theta \right)  \right) ^{
4}+2\,{B}^{2} \left(  \left( {Q}^{2}{r}^{2}-{r}^{4} \right) {B}^{2}+12
\,{Q}^{2}-4\,{r}^{2} \right)  \left( \cos \left( \theta \right) 
 \right) ^{2}+ \left( {B}^{2}{r}^{2}+4 \right) ^{2} \right] ^{6}{r}^{8
}}\, ,\label{eq:Kre_scalar}
\end{align}
\end{widetext}
where, $N(r,M,Q,B,\theta)$ refers to the numerator of the $\mathcal{K}$.
It is clear from \cref{eq:Ricci_sq,eq:Kre_scalar} that
quantities $R_{\mu\nu}R^{\mu\nu}$ and $\mathcal{K}$ contain two sorts of identical singularities, one is real at $r=0$, similar to the Schwarzschild spacetime, and the other is imaginary, which comes from the polynomial in $r$ in the denominator
}

\begin{figure*}
\begin{tabular}{c c}
\includegraphics[scale=0.65]{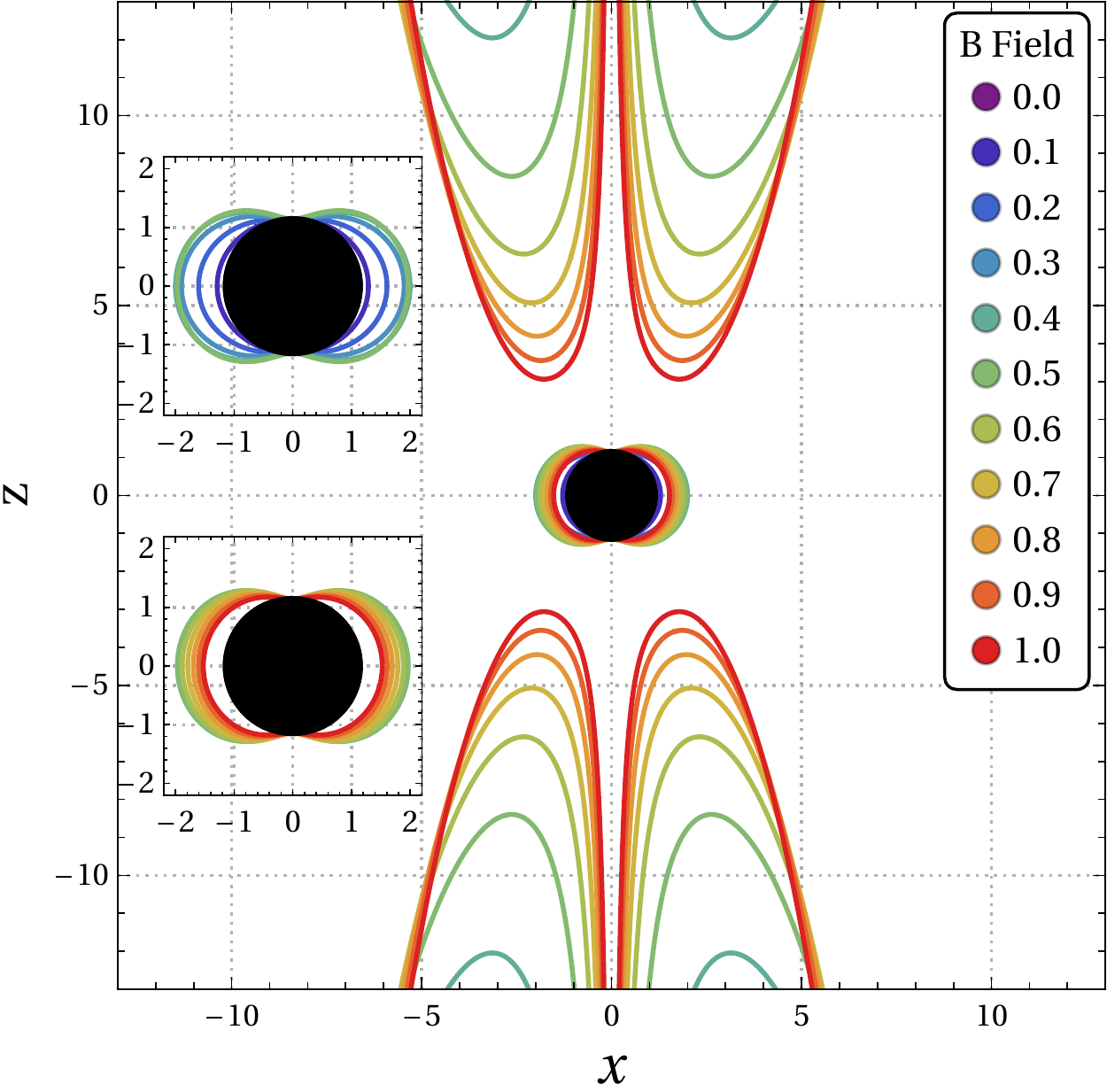}
&
\includegraphics[scale=0.65]{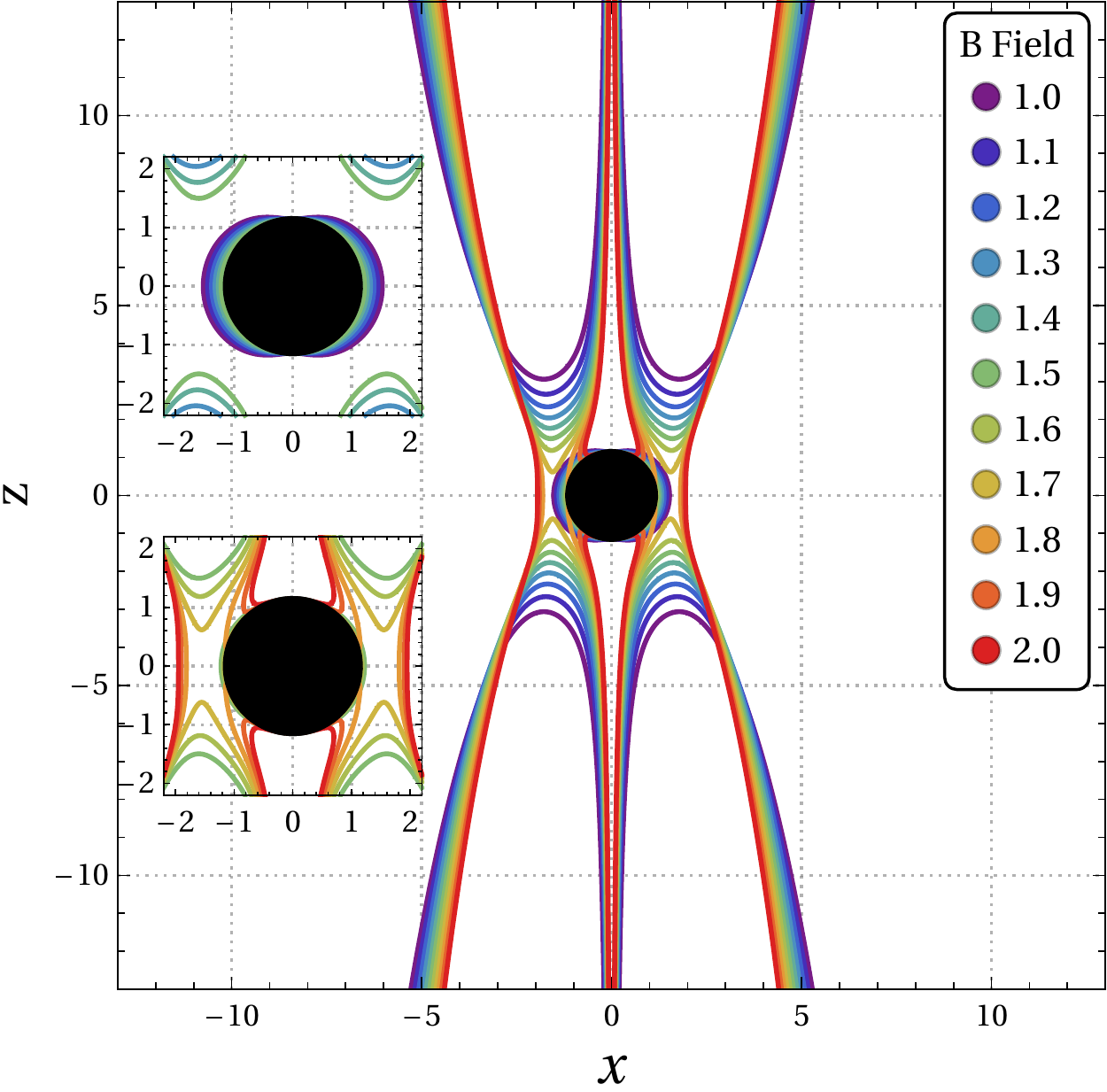}
\end{tabular}
\caption{\label{fig:ergo} For varying combinations of magnetic field $B$, the ergo region for the extremal axially symmetric magnetized black hole case is shown on the $x-z$ plane. In the {left panel} , the zoomed picture for cases when the magnetic field varies between $0$ and $0.5$ is shown in the {upper} inset plot. The {lower} inset plot, on the other hand, is a zoomed picture for cases when the magnetic field varies between $0.5$ and $1.0$. Similarly, in the {right panel}, the zoomed picture for cases when the magnetic field varies between $1.0$ and $1.5$ is shown in the {upper} inset plot. The {lower} inset plot, on the other hand, is a zoomed picture for cases when the magnetic field varies between $1.5$ and $2.0$. The charge and mass parameters of the black hole are set to $0.99$ and $1$, respectively.}
\end{figure*}

Due to stationary and axial symmetry, there are two Killing vectors in the magnetized black hole spacetime considered here and they are defined by 
$$\xi^{\mu}_{(t)}=\left(\frac{\partial}{\partial t}\right)^{\mu}\, \mbox{~~~and~~~}\\
\xi^{\mu}_{(\phi)}=\left(\frac{\partial}{\partial \phi}\right)^{\mu}\, ,
$$
The above two killing vectors, respectively, refer to a stationary and an axisymmetry of the spacetime, and thus there exist two conserved quantities, i.e., the energy and angular momentum for a charged particle along with its mass $m$. Let us then define the static limit surface SLS for which the Killing vector $\xi^{\mu}_{(t)} = \partial/\partial t$ becomes null, i.e., $g_{tt}= 0$.
We exhibit the variation of SLS as a function of $B$ field in the equatorial $\theta=\pi/2$ plane in \cref{fig:SLS1,fig:SLS2}.
 To extract the energy via Penrose process there must occur the ergosphere which exists in the region $r_{+}<r<r_{SLS}$, where $r_{SLS}$ refers to the static limit surface as the bounded stationary surface from outside. It can be defined by 
\begin{eqnarray}
\omega^2r^2\sin^2\theta-H^2F=0\, .
\end{eqnarray}
However, it turns out to be complicated to solve the above equation analytically. We therefore resort to its numerical evaluation to understand more deeply the property of the ergoregion around the magnetized RN black hole. In Fig.~\ref{fig:ergo}, we show the ergo region for the extremal axially symmetric magnetized black hole case on the plane $x-z$ for various combinations of magnetic field parameter $B$. As can be seen from Fig.~\ref{fig:ergo}, the ergo region can extend to infinity far away from black hole in both negative and positive $z$ directions. The point to be noted is that the ergoregion is separated from black hole for small values of magnetic parameter $B$, while it merges with the black hole for its large values as seen in {right panel} of \cref{fig:ergo}. It then results in increasing the appropriate volume of ergoregion for larger values of parameter $B$, thereby leading to arbitrarily high energy efficiency in the Penrose process. This is a remarkable nature of the magnetized black hole spacetime in contrast to other axially symmetric black holes.  
{It is worth noting here that in \cref{fig:ergo}, we consider only the extremal case (i.e. $Q=0.99 \approx 1$ and $M=1$) of the axially symmetric magnetized black hole and bring out the effect of magnetic field on the ergoregion in contrast with the case studied in \cite{Gibbons13} where non-extremal case (i.e. $Q<1$ and $M=1$) is discussed in detail. The main motivation to show the eroregion for the extremal case in \cref{fig:ergo} is coming from the well established fact in the literature that the energy extracted via Penrose process from the rotating black hole is  maximum only when the black hole satisfying the extremal condition (i.e. cauchy and event horizons coincide) \cite{1985JApA....6...85B}.}

The electromagnetic field around the magnetized black hole can be defined by 
\begin{eqnarray}\label{Eq:4-vec}
A &=& A_t dt + A_\phi (d\phi-\omega dt)\, , 
\end{eqnarray}
with the nonvanishing vector potential components (see details \cite{Shaymatov21c})  
 \begin{eqnarray}
A_t &=& -\frac{Q}{r} +\frac{3}{4} Q B^2 r\, (1+ F\cos^2\theta)\, ,\nonumber\\
A_\phi &=& \frac{2}{B} - H^{-1}\Big[\frac{2}{B} +
\frac{1}{2} B(r^2\sin^2\theta + 3 Q^2
\cos^2\theta)\Big]\, . \nonumber\\
\end{eqnarray}

\begin{figure}[ht]
\includegraphics[scale=0.6]{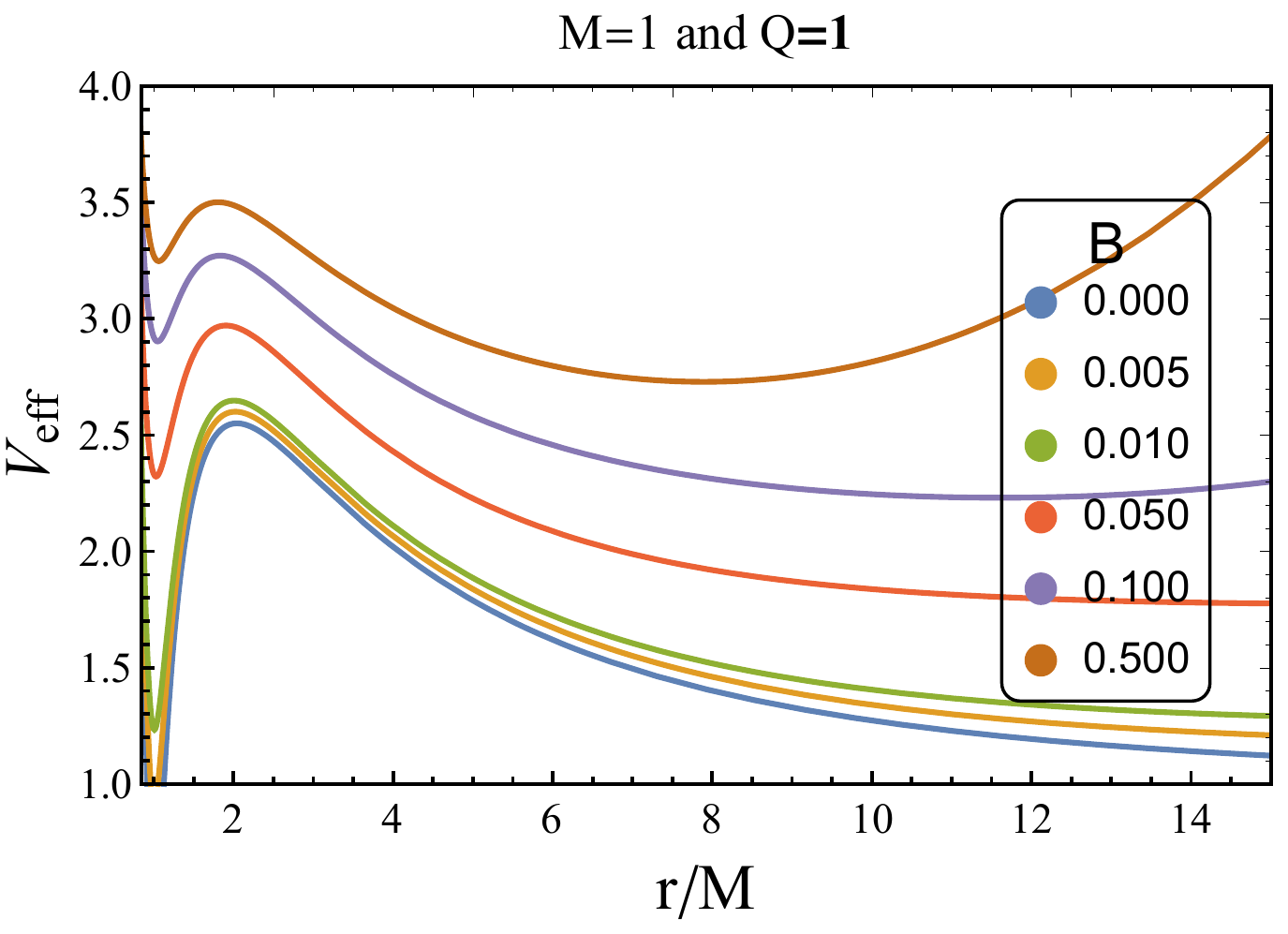}
\caption{\label{fig:eff_pot} Plot shows the effective potential as the function of $r/M$ in the equatorial plane, $\theta=\pi/2$ in the case with neutral particle. The effective potential $V_{eff}$ is plotted for various combinations of $B$ in the case of fixed $Q=1$. }
\end{figure}

\begin{figure*}
\begin{tabular}{c c }
  \includegraphics[scale=0.7]{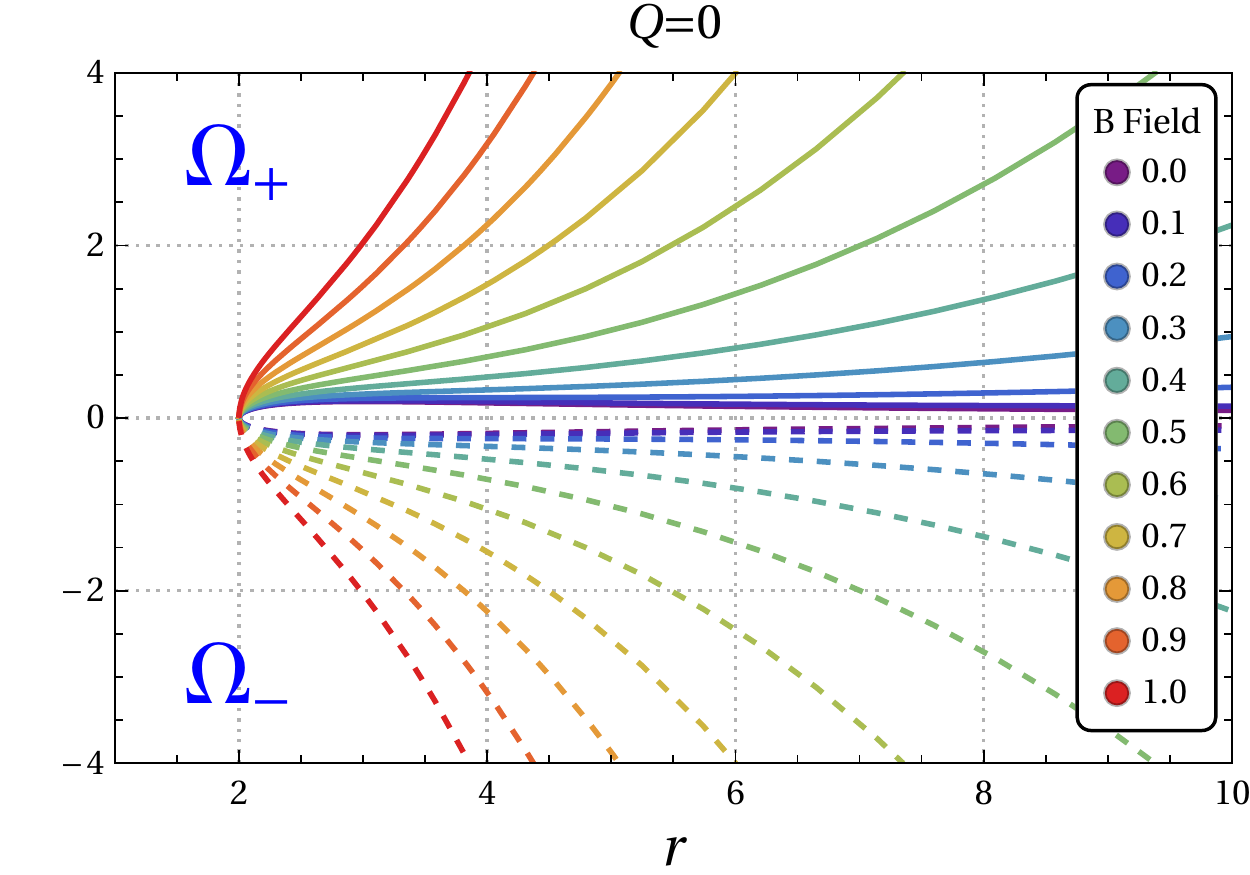}\hspace{-0.4cm}
  &  \includegraphics[scale=0.7]{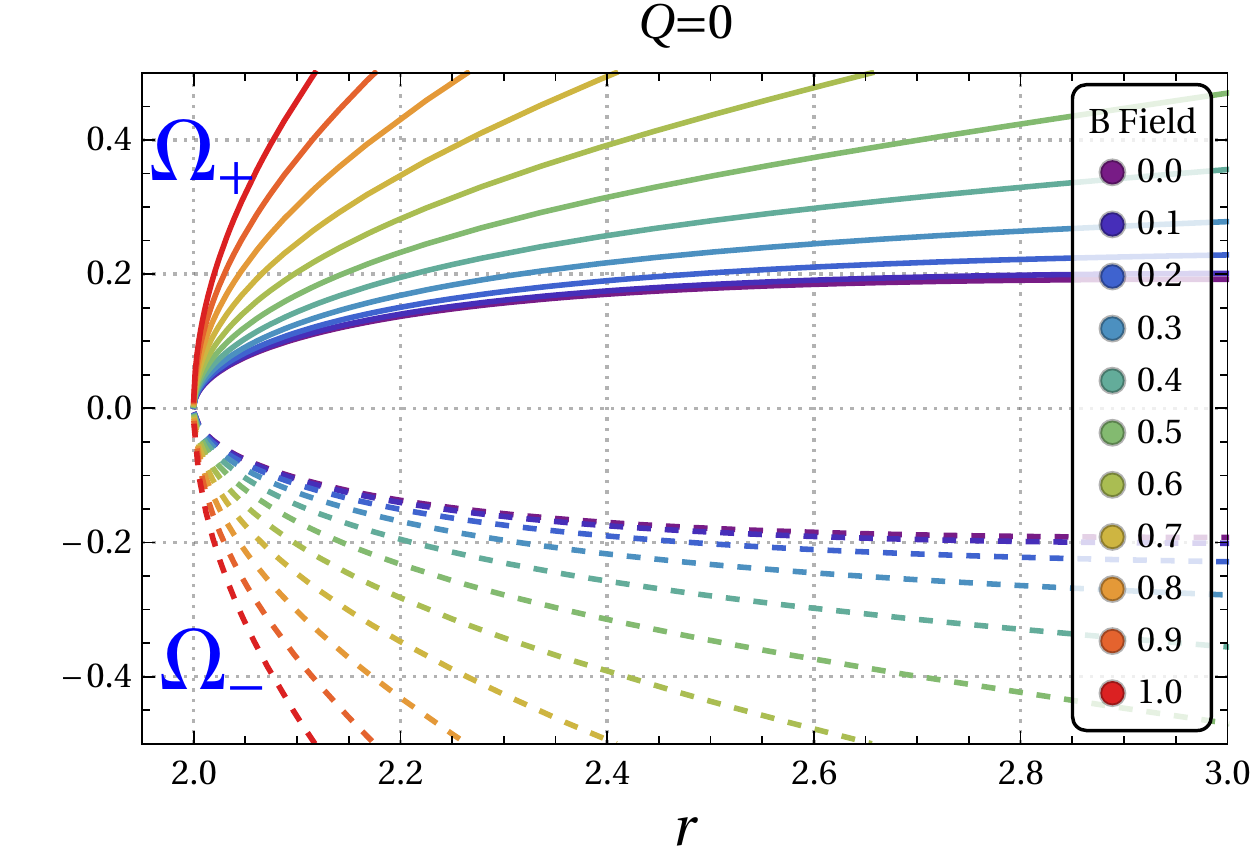}\\
  \includegraphics[scale=0.7]{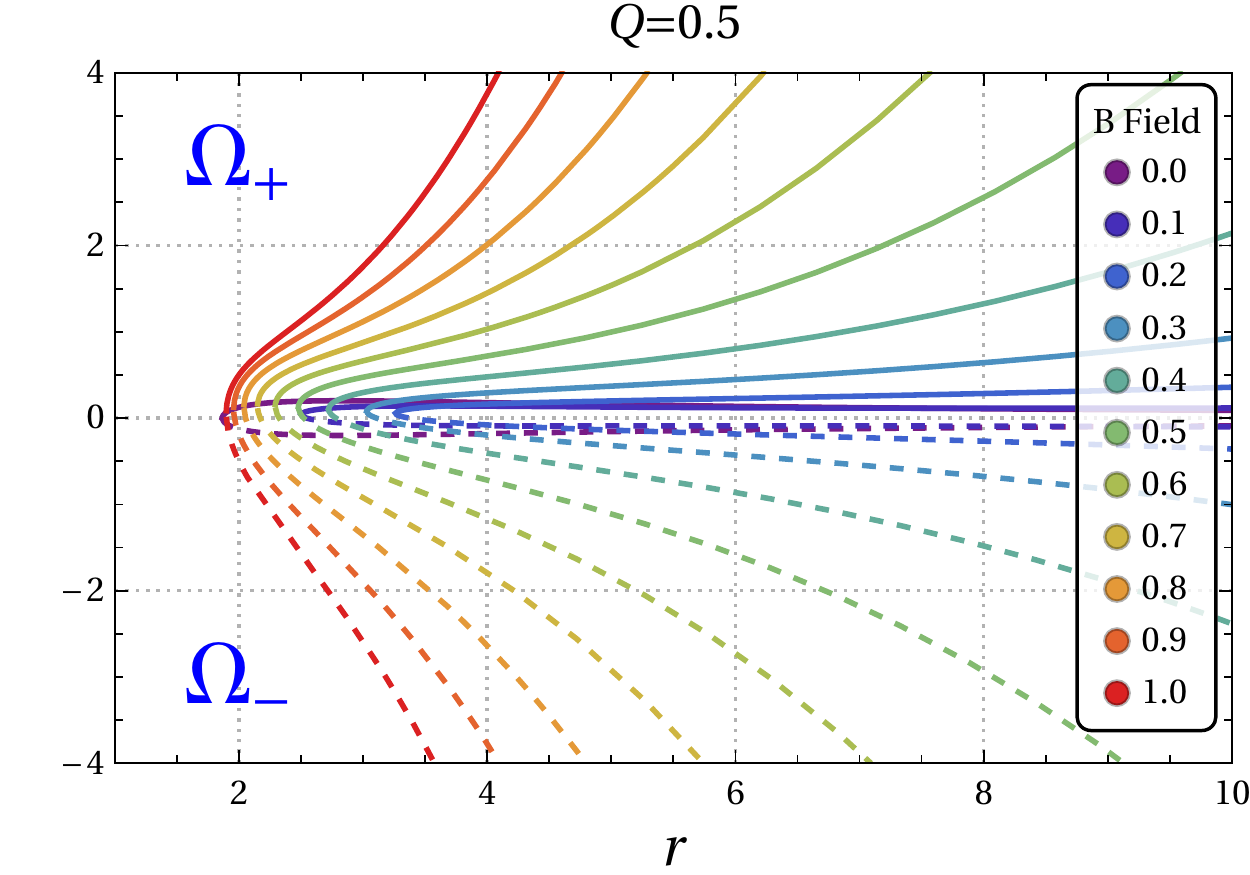}\hspace{-0.4cm}
  &  \includegraphics[scale=0.7]{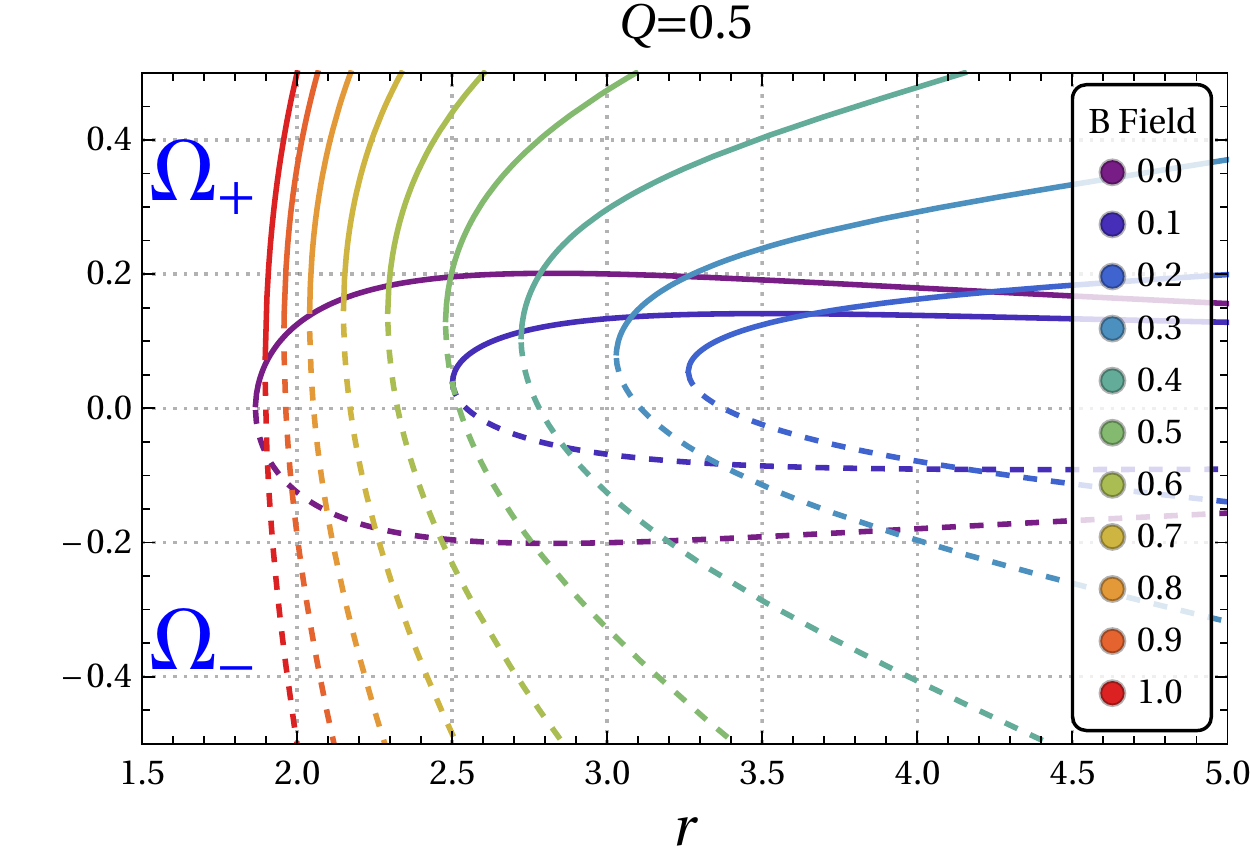}\\
    \includegraphics[scale=0.7]{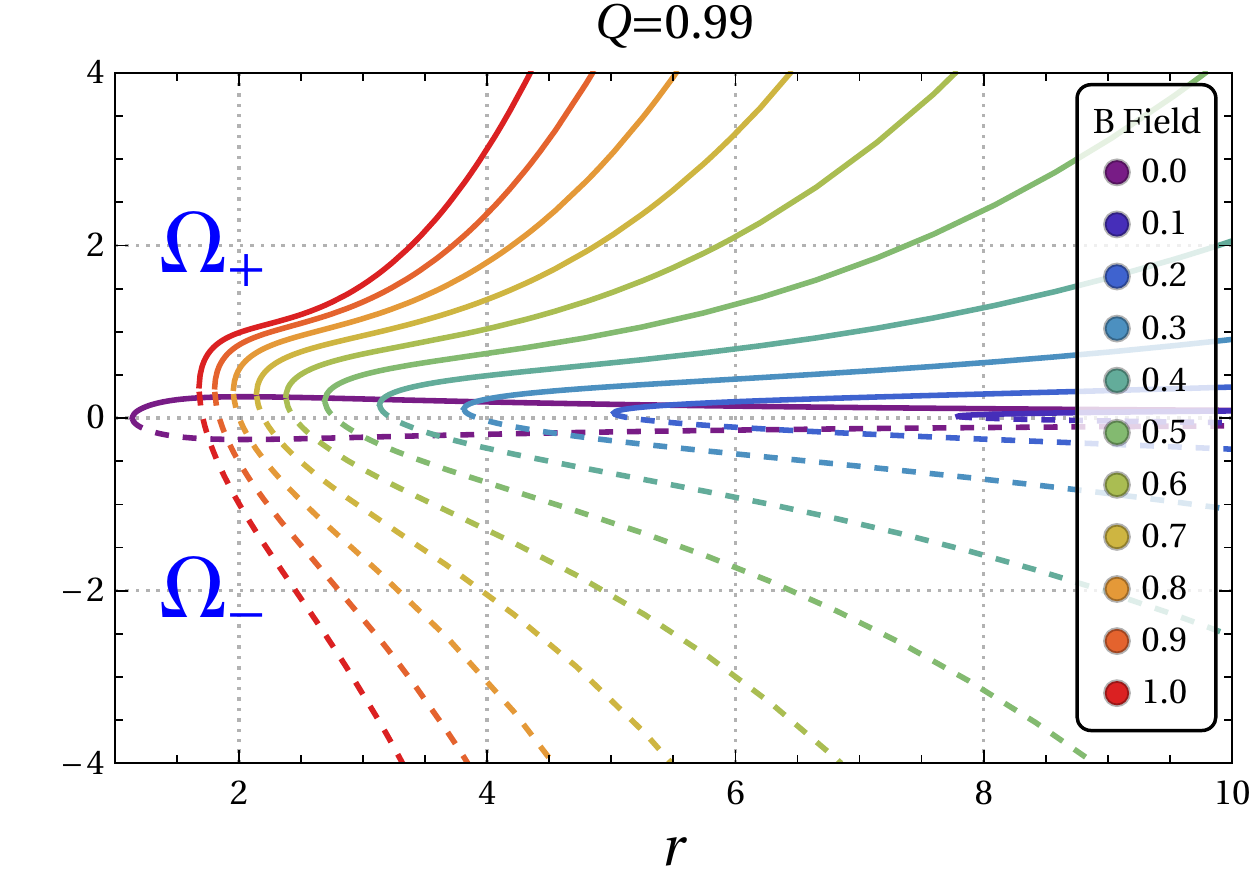}\hspace{-0.4cm}
  &  \includegraphics[scale=0.7]{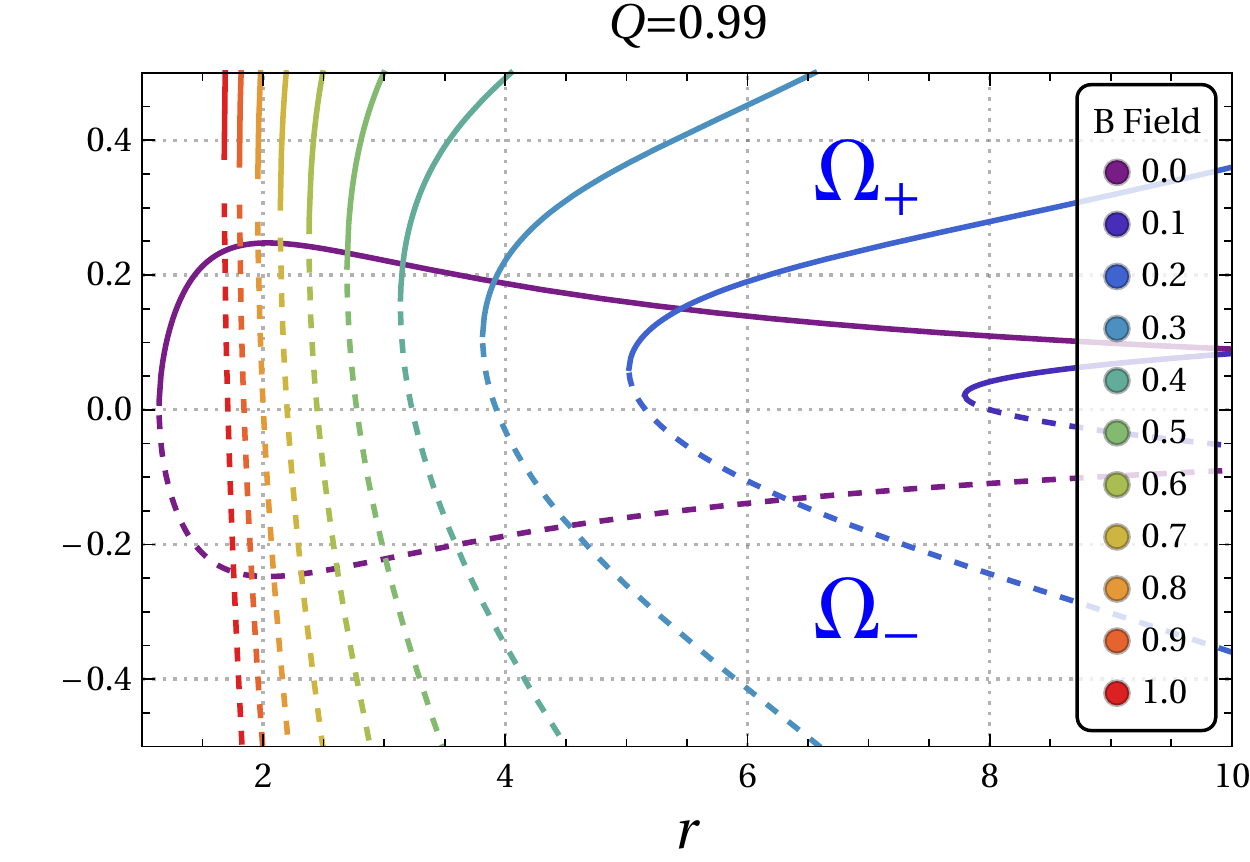}\\
\end{tabular}
	\caption{\label{fig:angular_vel} 
Plot shows the behaviour of angular velocity components $\Omega_{+}$ and $\Omega_{-}$ as a function of radial coordinate $r$ for various combinations of $Q$ and $B$. The {right column} is the zoomed out image of the {left column} for better visualization of the curves near to axes. The solid curves correspond to $\Omega_{+}$ case whereas the dotted curves correspond to $\Omega_{-}$ case, respectively. Here, we set the mass parameter $M=1$.}
\end{figure*}

\section{Particle dynamics} \label{Sec:Motion} 
Here, we explore a charged particle motion around axially symmetric magnetized black hole. For that we use the Hamiltonian which is defined by (see, for example~\cite{Misner73})
\begin{eqnarray}
H=\frac{1}{2}g^{\mu\nu} \left(\pi_{\mu}-qA_{\mu}\right)\left(\pi_{\nu}-qA_{\nu}\right)\, ,
\end{eqnarray}
with $\pi_{\mu}$ and $A_{\mu}$ which, respectively, correspond to the canonical momentum and electromagnetic four-vector potential. The charged particle's four momentum and the canonical momentum are related by the following relation 
\begin{eqnarray}
p^{\mu}=g^{\mu\nu}\left(\pi_{\nu}-qA_{\nu}\right)\, . 
\end{eqnarray}
For test particle motion one can write the equation of motion through the Hamiltonian as 
\begin{eqnarray} 
\label{Eq:eqh1}
  \frac{dx^\alpha}{d\lambda} = \frac{\partial H}{\partial \pi_\alpha}\,   \mbox{~~and~~}
  \frac{d\pi_\alpha}{d\lambda} = - \frac{\partial H}{\partial x^\alpha}\, , 
\end{eqnarray}
where $\lambda=\tau/m$ refers to affine parameter related by $\tau$ (i.e. the proper time for timelike geodesics). 

Imposing Eq.~(\ref{Eq:eqh1}) one can obtain the constants of motion for timelike geodesics: 
\begin{eqnarray}
\label{Eq:en} \pi_t-qA_{t}&=&
g_{tt}p^{t} + g_{t\phi}p^{\phi}\, ,\\
 \label{Eq:ln}
\pi_{\phi}-qA_{\phi}&=& g_{\phi t}p^{t} +
g_{\phi\phi}p^{\phi}\, .
\end{eqnarray}
On the basis of the normalization condition, $g_{\mu\nu}p^{\mu}p^{\nu}=-m^2$, along with Eqs.~(\ref{Eq:eqh1}-\ref{Eq:ln}), the timelike radial motion of the charged particle for the equatorial plane (i.e. $\theta=\pi/2$) can be defined by the effective potential  
\begin{eqnarray}
V_{eff}(r)&=&\frac{\Big[H q A_{\phi } \left(q A_{\phi }-2 \mathcal{L}\right)+H \mathcal{L}^2+r^2\Big]^{1/2}}{H^{1/2}\, r\, \Big[f H^2+r^2 (\omega -1) \omega \Big]^{-1/2} }\nonumber\\&-& q A_t+\omega  \left(\mathcal{L}-q A_{\phi }\right)\, ,
\end{eqnarray}
where we have defined $\mathcal{E}=E/m$ and
$\mathcal{L}=L/m$ for specific energy and angular momentum of the massive particle per unit mass. In Fig.~\ref{fig:eff_pot}, we show the radial dependence of the effective potential for neutral particle moving around the extremal axially symmetric magnetized black hole for various combinations of magnetic field parameter $B$. From Fig.~\ref{fig:eff_pot}, the effective potential grows as a function of $r/M$ as we increase the magnetic field parameter $B$, thus resulting in strengthening the potential barrier. It then results that a particle always starts its motion
in a distance not far away from black hole, as seen in Fig.~\ref{fig:eff_pot} (see for example \cite{Frolov10,Shaymatov21c}). 

As was mentioned above black hole rotation parameter causes the axially symmetric spacetime case, but the magnetized black hole solution may also generates the axially symmetric spacetime as a consequence of the existence of magnetic field parameter $B$. Further,
for the axially symmetric black hole case we know that to find the bounds on the allowed value of angular velocity, one needs to consider the circular motion of a test particle with $r=const$ and $\theta=const$. For this case ${\bf u}\sim {\bf \xi}_{(t)}+\Omega {\bf \xi}_{(\phi)}$, and  $\Omega=d\phi/dt=u^{\phi}/u^{t}$ is the angular velocity. The condition that the vector ${\bf u}$ is timelike requires $\Omega_{-}<\Omega<\Omega_{+}$, where
\begin{align}
  \Omega_{\pm} =\frac{-g_{t\phi}\pm \sqrt{(g_{t\phi})^{2}-g_{tt}g_{\phi\phi}}}{g_{\phi\phi}}.
\end{align}
The limiting values of $\Omega=\Omega_{\pm}$ correspond to the motion of photon. At the static limit surface ($g_{tt}=0$) $\Omega_{+}=0$ and  $\Omega_{-}=-2g_{t\phi}/g_{\phi\phi}$. However, outside the static limit surface $(g_{tt}<0)$,  $\Omega_{+}$ is always positive and  $\Omega_{-}$ is both positive and negative (see \cref{fig:SLS1,fig:SLS2,fig:angular_vel}). The four momentum of the falling particle 
\begin{eqnarray}\label{Eq:4-mom}
\pi_{\pm}=p^{t}(1,0,0,\Omega_{\pm})\, .
\end{eqnarray}
For the circular motion of a test particle with $r=const$ and $\theta=const$, $\upsilon_{(r,\theta)}=0$, one can write the following equation
\begin{eqnarray}\label{Eq:W0}
\left(g_{\phi\phi}\pi_t^2+g_{t\phi}^2\right)\Omega^2+2g_{t\phi}\left(\pi_t^2+g_{tt}\right)\Omega+g_{tt}\left(\pi_t^2+g_{tt}\right)=0\, ,\nonumber\\
\end{eqnarray}
with $\pi_t=-\left(\mathcal{E}+qA_{t}/m\right)$. 
From the above equation it is straightforward to obtain angular velocity for the infalling particle (see, for example~\cite{Parthasarathy86,Nozawa05})
\begin{eqnarray}\label{29}
\Omega=\frac{-g_{t\phi}\left(\pi_t^2+g_{tt}\right)+\sqrt{\left(\pi_t^2+g_{tt}\right)\left(g_{t\phi}^2-g_{tt}g_{\phi\phi}\right)\pi_t^2}}{g_{\phi\phi}\pi_t^2+g_{t\phi}^2}\, .\nonumber\\
\end{eqnarray}

\section{Energy extraction through Penrose process}\label{Sec:Penrose}

It is well known that recent modern astronomical observations show that the outflows that can have energies in the range of $E\approx 10^{42}-10^{47}\:\rm{erg/s}$ from
active galactic nuclei (AGN) in the form of winds and jets have been observed via x-ray, $\gamma$-ray and very long baseline interferometry (VLBI) observations~\cite{Fender04mnrs,Auchettl17ApJ,IceCube17b}. 
There is the relevance of the charged particle motion with these particle outflows coming out from AGN. In this regard, one needs to explore the Penrose process to propose an explanation for these observations. As mentioned earlier, this process was first theoretically described by Penrose in \cite{Penrose:1969pc}. An extensive analysis has since been developed in a large variety of situations \cite{Blandford1977,Wagh89,Morozova14,Alic12ApJ,Moesta12ApJ,Abdujabbarov11} addressing the effect of the magnetic field on the energy extraction mechanisms from black holes and accretion disks and jets in AGNs~\cite{McKinney07}. Also the effect of gravitomagnetic monopole charge on the high-energy phenomena has been considered 
\cite{Abdujabbarov11}.   

As per the Penrose mechanism, there must exist the ergoregion that is located outside the event horizon but inside the static radius so that energy can be extracted by a rotating black hole (i.e. the axially symmetric black hole spacetime). In this process, a massive particle falling into the ergosphere is forced to divide into two parts, and hence the momentum of these two pieces of particle allows one to escape to infinity from the black hole, while the other to enter the event horizon. As a result, the escaping one can have energy which is much greater than the one for an original piece of massive particle. It then turns out that black hole's energy is extracted as a consequence of the escaping piece of particle with a high energy. We then consider a massive particle falling into the ergosphere a far away from the black hole with energy $E_1\geq 1$. Let it be separated into two pieces with energies $E_{2}$ and $E_{3}$ in the ergoregion. As stated above, the escaping one attains high energy, $E_{3}>0$, while other one falls into the black hole with negative energy $E_{2}<0$. With this in view, one can write the conservation laws for particle parameters at the splitting point in the ergosphere of the black hole
\begin{eqnarray}\label{Eq:con_laws}
\begin{cases}
E_1=E_{2}+E_{3}\, \\
L_1=L_{2}+L_{3}\,  
\end{cases}
\mbox{~~~and~~~}
\begin{cases}
m_1=m_{2}+m_{3}\, \\
q_1=q_{2}+q_{3}\, . 
\end{cases}
\end{eqnarray}
where the condition, $E_{2}<0$ and $E_{3}\gg E_1$ is always satisfied as per the Penrose mechanism. 

According to the conservation law, one can write the four momentum at the point of split as follows \cite{1985JApA....6...85B,Nozawa05}
\begin{eqnarray}\label{Eq:con_law}
m_1u_1^{\mu}&=& m_2u_2^{\mu}+m_3u_3^{\mu}\, .
\end{eqnarray}
From the above conservation law, we consider $\phi$  component, i.e. $u^{\phi}=\Omega\, u^{t}=-\Omega B/A$, and thus Eq.~(\ref{Eq:con_law}) yields 
\begin{eqnarray}
\Omega_1m_1A_{1}B_2B_3=\Omega_2m_2A_{2}B_3B_1+\Omega_3m_3A_{3}B_2B_1\, ,
\end{eqnarray}
where we have defined $A=\mathcal{E} +qA_{t}/m$ and $B=g_{tt}+\Omega g_{t\phi}$. From the above equation we obtain 
\begin{eqnarray}
\frac{E_3+q_3A_{t}}{E_1+q_1A_{t}}=\left(\frac{\Omega_1B_2-\Omega_2B_1}{\Omega_3B_2-\Omega_2B_3}\right)\frac{B_3}{B_1}\, ,
\end{eqnarray}
and by algebraically manipulating the above expression the energy of the escaping particle can be written as
\begin{eqnarray}\label{Eq:E3}
E_3=\chi\left(E_1+q_1A_{t}\right)-q_3A_{t}\, .
\end{eqnarray}
Note that we have defined $\chi$ and $B_{i}$ as follows
\begin{eqnarray}\label{Eq:chi}
\chi=\left(\frac{\Omega_1-\Omega_2}{\Omega_3-\Omega_2}\right)\frac{B_3}{B_1}\, \mbox{~~and~~} B_{i}=g_{tt}+\Omega_{i}g_{t\phi}\, ,
\end{eqnarray}
where 
\begin{eqnarray}
\Omega_1= \Omega\, , \mbox{~~} \Omega_2=\Omega_{-}\, \mbox{~~and~~} \Omega_3=\Omega_{+}\, .
\end{eqnarray}
\begin{figure*}
\begin{tabular}{c c }
  \includegraphics[scale=0.60]{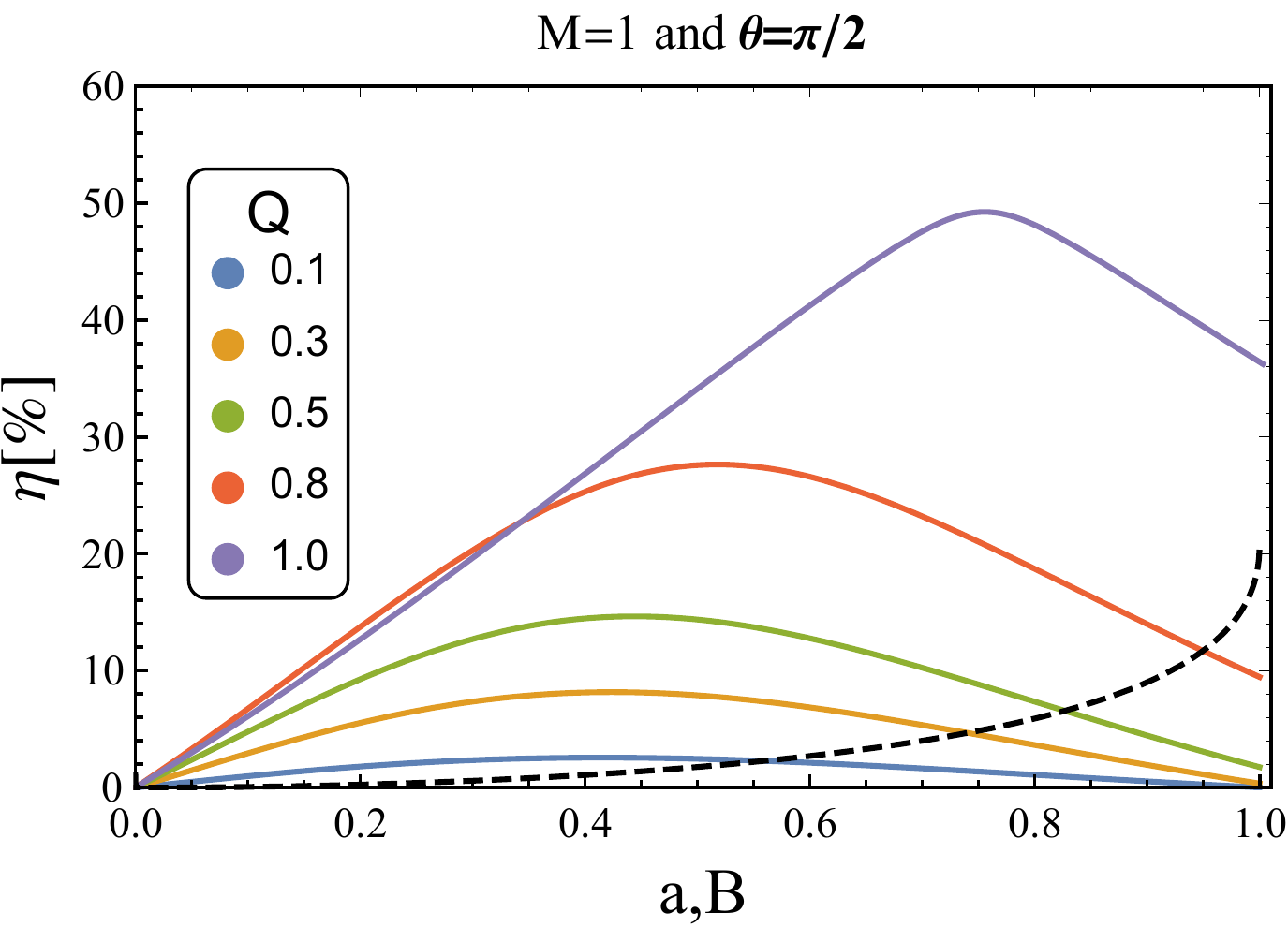}\hspace{-0.0cm}
  &  \includegraphics[scale=0.60]{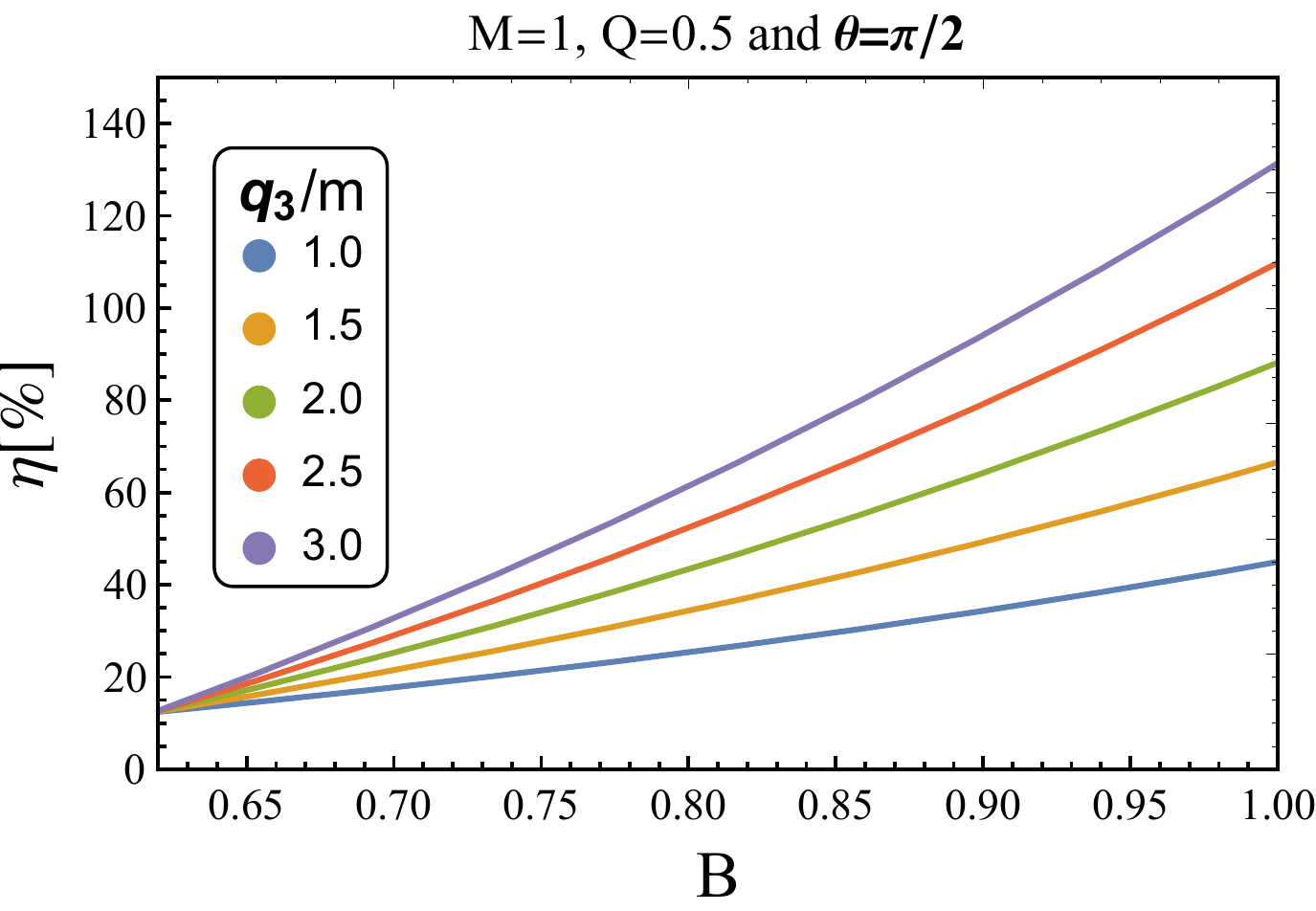}
\end{tabular}
\caption{\label{fig:en_eff} Plot shows the energy efficiency as the function of $B$ in the equatorial plane, $\theta=\pi/2$. Left panel: efficiency $\eta[\%]$ is plotted for various combinations of $Q$ in the case with neutral particle, i.e. $q_3=0$. Right panel: efficiency $\eta[\%]$ is plotted for various combinations of $q_{3}/m$ in the case of fixed $Q=0.5$. Note that the dashed line in the left panel shows the energy efficiency for rotating Kerr black hole. }
\end{figure*}

We then turn to discuss the energy efficiency that refers to the maximum energy extracted as the radiation due to the infalling matter into the ergoregion around the black hole. The energy efficiency via the Penrose process can be defined by following simple expression
\begin{eqnarray}
\eta= \frac{\vert E_2\vert}{E_1}=\frac{E_3-E_1}{E_1}\, .
\end{eqnarray}
By imposing Eqs.~(\ref{Eq:E3}) and (\ref{Eq:E3}), the energy efficiency for charged particles can be defined by
\begin{eqnarray}
\eta= \chi-1+\frac{q_3A_t}{m_1\pi_{t1}+q_1A_t}-\frac{ q_1A_t}{m_1\pi_{t1}+q_1A_t}\,\chi\, .
\end{eqnarray}
In the case of neutral particle, $q_1=0$, it is then separated into two charged
pieces with the conservation of charge, i.e. $q_2+q_3=0$. With this in view, the energy efficiency takes the form
as 
\begin{eqnarray}
\eta= \left(\frac{\Omega-\Omega_{-}}{\Omega_{+}-\Omega_{-}}\right)\left(\frac{g_{tt}+\Omega_{+}g_{t\phi}}{g_{tt}+\Omega\,g_{t\phi}}\right)-1-\frac{q_3A_t}{E_1}\, .
\end{eqnarray}
The above equation in the case of $q_3=0$ and $q_3\neq 0$ respectively reads as follows: 
\begin{eqnarray}\label{Eq:q=0}
\eta\vert_{q_3=0}&=& \frac{1}{2\left(4+B^2 \left(1+\sqrt{1-Q^2}\right)^2\right)} \nonumber\\&\times&\left[\left(-8 B Q \left(1+\sqrt{1-Q^2}\right) \right.\right. \nonumber\\ &\times&\left(4-B^2 \left(1+\sqrt{1-Q^2}\right)^2\right)\nonumber\\&+&\left.\left(4+B^2 \left(1+\sqrt{1-Q^2}\right)^2\right)^2\right)^{1/2}\nonumber\\&-&\left.\left(4+B^2 \left(1+\sqrt{1-Q^2}\right)^2\right)\right]\, ,
\end{eqnarray}
and
\begin{eqnarray}\label{Eq:q}
\eta&=& \eta\vert_{q_3\neq0} -\frac{q_3}{E_1}\left[-\frac{Q}{1+\sqrt{1-Q^2}}\right.\nonumber\\&+&\left.\frac{3}{4}QB^2\left(1+\sqrt{1-Q^2}\right)\right] \, .
\end{eqnarray}
For further analysis we shall for simplicity consider $E_1/m_1=1$. {In doing so, we can further assume that $q_3/E_1=q/m$ at the splitting point occurring around black hole,  especially very close to the black hole's horizon $r = r_+$. However, the location of the splitting point can also exist at a large distance from the black hole's horizon as the ergo region extends; see Fig.~\ref{fig:ergo}. In order to be more precise we shall restrict ourselves to the case for which the particle's splitting point occurs in the close vicinity of the back hole's horizon. }

We now analyze the energy efficiency, $\eta$, as the function of $B$ for various combinations.  In Fig.~\ref{fig:en_eff}, we show the energy efficiency extracted from the magnetized black hole as that of the infalling mater into the ergoregion. The left panel shows the impact of the combined effect of black hole charge $Q$ and magnetic field parameter $B$ on the energy efficiency, while the right panel shows the same behavior for fixed $Q=0.5$ in the case with the charged particle that refers to the escaping piece of massive particle falling into the ergosphere.  As shown in Fig.~\ref{fig:en_eff} (left panel), the shape of the energy efficiency shifts up to higher $\eta$ with increasing $Q$. However, it slightly gets decreased as the magnetic field parameter $B$ increases. This happens because the area of the ergosphere required for energy extraction turns out to be getting smaller beyond $B>B_{cr}$; see Fig~\ref{fig:ergo}. It is worth noting that the maximum value of the efficiency reaches up to the value greater than 50 $\%$ which is comparable value for Kerr black hole case. This is a remarkable property of the axially symmetric magnetized Reissner-Nordstr\"{o}m black hole. As stated above we show the energy efficiency for neutral particle, i.e. $q_3=0$ in the left panel, whereas in the right panel we show that energy efficiency can exceed 100 $\%$ in the case of charged particle $q_3\neq 0$. This happens because the second term of Eq.~(\ref{Eq:q}) gives rise to the main contribution so that $\eta$ reaches its large values. This contribution stems from the combined effects of black hole charge and magnetic field on the charged particle that escapes from the black hole. Another interesting reason for that the parameter, $q/m$,  starts to take larger values as a charged particle gets elementary charged one. Also we would like to underline that since the process with large $q/m>1$ for astrophysical black holes might be relevant to astrophysics we can take overestimated values of the parameter $q/m$, and thus we observed that the energy efficiency can exceed over 100 $\%$ in the case of charged particle escaping with the arbitrarily high energy from the axially symmetric magnetized Reissner-Nordstr\"{o}m black hole.

\section{Conclusions}
\label{Sec:conclusion}
We demonstrated that the Penrose process can also be applied to the axially symmetric magnetized Reissner-Nordst\"{o}m black hole, which is in contrast to previous studies that have only looked at the rotating black holes. We started with the investigation of the ergosphere and bring out the effect of the electric charge and magnetic field on it. We found that the static limit surface SLS first increases till the magnetic field parameter reaches $B=0.4$ and then begins to decrease again as the $B$-field increases further for the fixed value of the electric charge $Q$ (see \cref{fig:SLS1}). Contrary to this, the SLS always decreases as $Q$ increases for the fixed value of $B$-field parameter as shown in \cref{fig:SLS2}.

The ergoregion around the axially symmetric magnetized Reissner-Nordstr\"{o}m black hole is discovered to be caused solely by the magnetic field, which favors the formation of negative-energy states.  The presence of a magnetic field has resulted in two interesting findings: first, it allows one of the particles to have a higher negative energy after splitting in the ergoregion, and second, negative energy states can also exist far away from the black hole's event horizon in the form of potential wells. It is also noted that when the magnetic field increases, the depth of these potential wells grows, eventually merging into the ergosphere (see \cref{fig:ergo}). The former observation led to increased Penrose process efficiency for both neutral (see {left panel} of \cref{fig:en_eff}) and charged particles, as shown in the {right panel} of \cref{fig:en_eff}, whereas the latter observation brought the region of negative energy states within easy reach of a far away observer. It is interestingly observed that for the case of neutral particle the efficiency of the Penrose process for the extremal ($Q=M$) case of the axially symmetric magnetized Reissner-Nordst\"{o}m is more than double ($\approx 50\%$) to that of the extremal Kerr ($a=M$) black hole case which is $\approx 20\%$, as shown in {left panel} of \cref{fig:en_eff}.  Furthermore, it is observed that for the case neutral particle, the efficiency first increases, reaches a maximum value at a specific value of $B$, and then begins to decrease again as B increases, as for all values of $Q$ in the range $0\to 1$, in contrast to the Kerr black hole case, where the efficiency increases continuously as the spin of the black hole increases. 
The magnetized Reissner-Nordst\"{o}m black hole solution generates the axially symmetric spacetime similarly to what is observed in vicinity of rotating Kerr black hole. From this point of view, the magnetic field parameter can be regarded as equivalent to the spin parameter of Kerr black hole. One can easily notice that the axially symmetric magnetized Reissner-Nordst\"{o}m and Kerr black hole geometries can have similar energy extraction efficiencies, thereby resulting in mimicking the black hole spin up to $a=1$ by the magnetic parameter of black hole. From an astrophysical point of view, a distant observer is not able to distinguish two geometries and thus radiations emitted by accretion disk in low mass x-ray binaries may provide similar observations due to the above mentioned black hole spacetimes degeneracy. The study explored here has however relevance to an idealized theoretical model, it does play increasingly important role to understand more deeply the novel and qualitative aspects of the axially symmetric magnetized Reissner-Nordst\"{o}m black hole spacetime.   

On the other hand, when the charged particle scenario (i.e. $q_{3}/m\neq 0$) is analyzed for $Q=0.5$, the efficiency increases continuously as $B$ expands, as shown in the {right panel} of \cref{fig:en_eff}. When this figure is examined further, it is found that in the extreme case (i.e., when $B=1$), the efficiency of the Penrose process is always larger than that of the Kerr black hole case when $q_{3}/m\geq1$. Additionally, when $q_{3}/m\geq2.2$, the efficiency becomes more than $100\%$. Nonetheless, a similar observation is reported in \cite{Dadhich:2018gmh}, where the authors examine the Kerr black hole immersed in an external magnetic field.

It is also important to explore the role of the magnetic field in the vicinity of a black hole from an astrophysical viewpoint. Some fascinating study has already pointed out that the magnetic field's impact on the accretion disc is very obvious \cite{Blandford1977,PhysRevD.17.1518}.


\begin{acknowledgements}
P. S. acknowledges support under University Grant Commission (UGC)-DSKPDF scheme (Govt. of India) through grant No. F.4-2/2006(BSR)/PH/20-21/0053. He would also like to thank the IUCAA Centre for Astronomy Research and Development (ICARD), Gurukula Kangri (Deemed to be University), Haridwar, India, for the facilities utilized. S.S. and B.A. acknowledge the support from Research F-FA-2021-432 of the Uzbekistan Ministry for Innovative Development. 
\end{acknowledgements}

%
\bibliographystyle{apsrev4-1}  
\bibliography{MBH_Pen_P}

\end{document}